\documentclass[conference,a4paper]{IEEEtran}
\IEEEoverridecommandlockouts{}

\newif\ifarxiv
\newif\ifedas
\newif\ifdraft

\edasfalse
\arxivtrue
\draftfalse

\ifedas
    \ifarxiv
        \PackageError{conditionals.tex}{Cannot set both 'edas' and 'arxiv' to true}{}%
    \fi 
\fi

\usepackage{cite}
\usepackage{amsmath,amssymb,amsfonts}
\usepackage{algorithm}
\usepackage{algorithmic}
\usepackage{subcaption}
\usepackage{graphicx}
\usepackage{textcomp}
\usepackage{xcolor}
\ifedas
\else
	\usepackage{orcidlink} 
\fi
\usepackage{xstring} 
\usepackage{xifthen}
\usepackage{mfirstuc}
\usepackage{tikz}
\usepackage{tikzscale}
\usepackage{psfrag}
\usepackage{type1cm }
\usepackage{booktabs}

\usetikzlibrary{decorations.pathreplacing,calligraphy}

\ifedas
    \newcommand{\myRef}[3]{#1~\ref*{#2:#3}} 
    \newcommand{\myRefLong}[3]{#1~\ref*{#2:#3} - \nameref*{#2:#3}} 
\else
    \usepackage{hyperref}
    \hypersetup{colorlinks=false}

    \newcommand{\myRef}[3]{\hyperref[{#2:#3}]{#1~\ref*{#2:#3}}} 
    \newcommand{\myRefLong}[3]{\hyperref[{#2:#3}]{#1~\ref*{#2:#3} - \nameref*{#2:#3}}} 
\fi

\newcommand{\gleichung}[1]{(\ref{eq:#1})\xspace}

\newcommand{\Abbildung}[1]    {\myRef{Fig.}     {fig}  {#1}\xspace}
\newcommand{\abbildung}[1]    {\myRef{Fig.}     {fig}  {#1}\xspace}

\newcommand{\Abschnitt}[1]    {\myRef{Section}  {sec}  {#1}\xspace}
\newcommand{\abschnitt}[1]    {\myRef{Section}  {sec}  {#1}\xspace}

\newcommand{\Tabelle}[1]    {\myRef{Table}  {tab}  {#1}\xspace}
\newcommand{\tabelle}[1]    {\myRef{Table}  {tab}  {#1}\xspace}

\def\BibTeX{{\rm B\kern-.05em{\sc i\kern-.025em b}\kern-.08em
    T\kern-.1667em\lower.7ex\hbox{E}\kern-.125emX}}

\ifdraft
  \usepackage{xcolor}
  \usepackage{graphicx}
  \usepackage{eso-pic}

  \definecolor{draftred}{RGB}{255,200,200} 

  \AddToShipoutPictureBG{%
    \AtPageCenter{%
      \makebox(0,0){%
        \rotatebox{45}{%
          \textcolor{draftred}{%
            \fontsize{8cm}{8cm}\selectfont
            \bfseries DRAFT%
          }%
        }%
      }%
    }%
  }
\fi
\usepackage{xspace}
\usepackage{amsmath}
\usepackage{amsfonts}
\usepackage{amssymb}
\usepackage{bm}
\usepackage{upgreek}
\usepackage{graphicx}
\usepackage{ifthen}
\usepackage[dvipsnames]{xcolor}

\definecolor{tuiblue}{HTML}{003359}
\definecolor{tuigreen}{HTML}{00747A}
\definecolor{tuiorange}{HTML}{FF7900}
\definecolor{tuigrey}{HTML}{A5A5A5}

\definecolor{tuiblueCMYK}{cmyk}{1,0.7,0.1,0.5}
\definecolor{tuigreen}{cmyk}{1,0,0.45,0.2}
\definecolor{tuiorange}{cmyk}{0,0.6,1,0}

\newcommand{\setN}{\ensuremath{\mathbb{N}}}  
\newcommand{\setR}{\ensuremath{\mathbb{R}}}  
\newcommand{\setC}{\ensuremath{\mathbb{C}}}  

\renewcommand{\vec}[1]{\ensuremath{\bm{#1}}}

\newcommand{\mat}[1]{\vec{#1}}

\newcommand{\func}[1]{\ensuremath{\mathrm{#1}}}                          

\newcommand{\mTr}{\ensuremath{^\mathrm{T}}}           

\newcommand{\mInv}{\ensuremath{^{\text{-}1}}}                

\newcommand{\mNorm}[2]{\ensuremath{\left\| #1 \right\|_{#2}}}         
\newcommand{\mFrobNorm}[1]{\mNorm{#1}{\text{F}}}                      

\makeatletter

\newcommand{\circConvolve}{\ensuremath{\mathbin{\mathpalette\make@circled\ast}}}
\newcommand{\make@circled}[2]{%
  \ooalign{$\m@th#1\smallbigcirc{#1}$\cr\hidewidth$\m@th#1#2$\hidewidth\cr}%
}
\newcommand{\smallbigcirc}[1]{%
  \vcenter{\hbox{\scalebox{0.77778}{$\m@th#1\bigcirc$}}}%
}
\makeatother

\newcommand{\hslashslash}{%
  \raisebox{.9ex}{%
    \scalebox{.7}{%
      \rotatebox[origin=c]{18}{$-$}%
    }%
  }%
}

\DeclareRobustCommand{\coele}{%
  {
    \ensuremath{%
    \vphantom{\vartheta}%
    \ooalign{\kern-.07em\smash{\lower.35em\hbox{\hslashslash}}\hidewidth\cr$\vartheta$\cr}%
    \kern.05em
    }%
  }%
}

\newcommand{\upcoele}{%
  {
    \text{
      \ensuremath{%
        \vphantom{\upvartheta}%
        \ooalign{\kern-.07em\smash{\lower.35em\hbox{\hslashslash}}\hidewidth\cr$\upvartheta$\cr}%
        \kern.05em%
      }%
    }%
  }%
}

\newcommand{\diff}{\mathrm{d}}

\let\oldpartial\partial
\DeclareRobustCommand{\uppartial}{\text{\rotatebox[origin=t]{20}{\scalebox{0.95}[1]{$\oldpartial$}}}\hspace{-1pt}}
\renewcommand{\partial}{\uppartial}

\newcommand{\pDiff}{\uppartial} 

\newcommand{\deriv}[3][]{%
  \ifthenelse{\equal {#1}{}}{
    \frac{\diff #2}{\diff #3}%
  }{ 
    \frac{\diff^{#1} #2}{\diff #3^{#1}}%
  }%
}

\newcommand{\pDeriv}[3][]{%
  \ifthenelse{\equal {#1}{}}{
    \frac{\pDiff #2}{\partial #3}%
  }{
    \frac{\pDiff^{#1} #2}{\partial #3^{#1}}%
  }%
}

\newcommand{\mathTransform}[4]{%
  \ensuremath{%
      \operatorname*{#2}%
      \ifthenelse{\equal {#4}{}}{
      }{
        _{#4} %
      }%
      \kern-2.0pt%
      \left\lbrace #3 \right\rbrace%
      \ifthenelse{\equal {#1}{}}{
      }{
        \kern-2.0pt #1 %
      }%
  }%
}

\newcommand{\mathContTransform}[4]{%
  \mathTransform%
  { \if\relax\detokenize{#1}\relax\else\ensuremath{\left( #1 \right)}\fi }
  {#2}{#3}{#4}%
}

\newcommand{\mathDiscTransform}[4]{%
  \mathTransform%
  { \if\relax\detokenize{#1}\relax\else\ensuremath{\left[ #1 \right]}\fi }
  {#2}{#3}{#4}%
}

\newcommand{\mathContInvTransform}[4]{\mathContTransform{#1}{#2\mInv}{#3}{#4}}

\newcommand{\mathDiscInvTransform}[4]{\mathDiscTransform{#1}{#2\mInv}{#3}{#4}}

\makeatletter
\newcommand{\transformRelation}[1][]{%
  \mathop{%
    \mbox{%
      \setlength{\unitlength}{0.1em}%
      \begin{picture}(37,10)%
        \put(4,3){\circle{4}}%
        \put(6,3){\line(1,0){26}}%
        \put(33,3){\circle*{4}}%
        \put(19,7){\makebox(0,0){#1}}%
      \end{picture}%
    }%
  }\slimits@%
}
\makeatother

\makeatletter
\newcommand{\revTransformRelation}[1][]{%
  \mathop{%
    \mbox{%
      \setlength{\unitlength}{0.1em}%
      \begin{picture}(37,10)%
        \put(4,3){\circle*{4}}%
        \put(6,3){\line(1,0){25}}%
        \put(33,3){\circle{4}}%
        \put(19,7){\makebox(0,0){#1}}%
      \end{picture}%
    }%
  }\slimits@%
}
\makeatother

\newcommand{\createContinuousTransform}[2]{%
  \expandafter\NewDocumentCommand\csname#1Transform\endcsname{E{_}{\relax}O{}m}{%
    \mathContTransform{##2}{#2}{##3}{##1}%
  }%
  \expandafter\NewDocumentCommand\csname#1InvTransform\endcsname{E{_}{\relax}O{}m}{%
    \mathContInvTransform{##2}{#2}{##3}{##1}%
  }%
  \expandafter\newcommand\csname#1Relation\endcsname{%
    \mathop{\transformRelation[#2]{}}
  }%
  \expandafter\newcommand\csname#1RevRelation\endcsname{%
    \mathop{\revTransformRelation[#2]{}}
  }%
}

\newcommand{\createDiscreteTransform}[2]{%
  \expandafter\NewDocumentCommand\csname#1Transform\endcsname{E{_}{\relax}O{}m}{%
    \mathDiscTransform{##2}{#2}{##3}{##1}%
  }%
  \expandafter\NewDocumentCommand\csname#1InvTransform\endcsname{E{_}{\relax}O{}m}{%
    \mathDiscInvTransform{##2}{#2}{##3}{##1}%
  }%
  \expandafter\newcommand\csname#1Relation\endcsname{%
    \mathop{\transformRelation[#2]{}}
  }%
  \expandafter\newcommand\csname#1RevRelation\endcsname{%
    \mathop{\revTransformRelation[#2]{}}
  }%
}

\newcommand{\fourierLetter}{\ensuremath{\mathcal{F}}}

\createContinuousTransform{fourier}{\fourierLetter}

\newcommand{\dftLetters}{\ensuremath{\mathcal{DFT}}}

\createDiscreteTransform{dft}{\dftLetters}

\newcommand{\laplaceLetter}{\ensuremath{\mathcal{L}}}

\createContinuousTransform{laplace}{\laplaceLetter}

\NewDocumentCommand\expValue{e{_}m}{%
    \operatorname*{\mathbb{E}}\IfNoValueF{#1}{_{#1}} \left\lbrace #2 \right\rbrace%
}

\NewDocumentCommand\variance{e{_}m}{%
    \operatorname*{\mathrm{var}}\IfNoValueF{#1}{_{#1}} \left\lbrace #2 \right\rbrace%
}

\NewDocumentCommand\covariance{e{_}m}{%
    \operatorname*{\mathrm{cov}}\IfNoValueF{#1}{_{#1}} \left\lbrace #2 \right\rbrace%
}

\NewDocumentCommand\correlation{e{_}m}{%
    \operatorname*{\mathrm{corr}}\IfNoValueF{#1}{_{#1}} \left\lbrace #2 \right\rbrace%
}

\usepackage{amsmath,amssymb,amsfonts}
\usepackage{xspace}

\DeclareMathOperator*{\argmin}{\func{arg}\,\func{min}}

\newcommand{\Label}[1]{{\scriptscriptstyle\mathrm{#1}}}

\newcommand{\SamplingFrequency}{\ensuremath{f_\Label{S}}\xspace}

\newcommand{\BasePulse}[3]{\ensuremath{\func{s}_{#1,#2}\!\left(#3\right)}\xspace}

\newcommand{\TxSignal}[2]{\ensuremath{\func{x}_{#1}\!\left(#2\right)}\xspace}

\newcommand{\PulseStartTime}[2]{\ensuremath{\tau_{#1}\!\left[#2\right]}\xspace}

\newcommand{\FirstPulseStartTime}[1]{\ensuremath{\tau_{\Label{start}, #1}}\xspace}

\newcommand{\RxTxPulseTimeDiff}[3]{\ensuremath{\tau_{\Label{\Delta},#1,#2}\!\left(#3\right)}\xspace}

\newcommand{\PulseDuration}[2]{\ensuremath{T_{#1}\!\left[#2\right]}\xspace}

\newcommand{\FirstPulseDuration}[1]{\ensuremath{T_{\Label{start}, #1}}\xspace}

\newcommand{\FirstPulseFrequency}[1]{\ensuremath{f_{\Label{start}, #1}}\xspace}

\newcommand{\EstimatedFirstPulseDuration}[2]{\ensuremath{\hat{T}_{\Label{start}, #1,#2}}\xspace}

\newcommand{\PropagationDuration}[4]{\ensuremath{\delta_{#1,#2,#3}}}

\newcommand{\TxPulseIndex}[2]{\ensuremath{n_{\Label{T},#1}\ifstrempty{#2}{}{\!\left(#2\right)}}}

\newcommand{\RxPulseIndex}[3]{\ensuremath{n_{\Label{R},#1,#2}\ifstrempty{#3}{}{\!\left(#3\right)}}}

\newcommand{\RxTxPulseIndexDiff}[3]{\ensuremath{n_{\Label{\Delta},#1,#2}\!\left(#3\right)}}

\newcommand{\TxPhase}[2]{\ensuremath{\func{\varphi_{\Label{T},#1}\!\left(#2\right)}}}

\newcommand{\RxPhase}[3]{\ensuremath{\func{\varphi_{\Label{R},#1,#2}\!\left(#3\right)}}}

\newcommand{\RealRxPhaseDiff}[3]{\ensuremath{\tilde{\func{\psi}}_{#1,#2}\!\left(#3\right)}}

\newcommand{\RxPhaseDiff}[3]{\ensuremath{\func{\psi}_{#1,#2}\!\left(#3\right)}}

\newcommand{\RxPhaseDiffMat}{\ensuremath{\mat{\varPsi}}\xspace}

\newcommand{\EstimatedRxPhaseDiffMat}[1]{\ensuremath{\mat{\hat{\varPsi}}_{#1}}\xspace}

\newcommand{\EstimatedRxPhaseDiffVec}[1]{\ensuremath{\vec{\hat{\psi}}_{#1}}\xspace}

\newcommand{\PulseStartTimeDepRxPhaseDiff}[3]{\ensuremath{\func{\psi}_{\Label{\tau},#1,#2}\ifstrempty{#3}{}{\!\left(#3\right)}}}

\newcommand{\PulseStartTimeDepRxPhaseDiffMat}{\ensuremath{\mat{\varPsi}_{\Label{\tau}}}\xspace}

\newcommand{\EstimatedPulseStartTimeDepRxPhaseDiffMat}[1]{\ensuremath{\mat{\hat{\varPsi}}_{\Label{\tau}, #1}}\xspace}

\newcommand{\PropagationTimeDepRxPhaseDiff}[3]{\ensuremath{\func{\psi}_{\Label{\delta},#1,#2}\!\left(#3\right)}}

\newcommand{\PropagationTimeDepRxPhaseDiffMat}{\ensuremath{\mat{\varPsi}_{\Label{\delta}}}\xspace}

\newcommand{\EstimatedPropagationTimeDepRxPhaseDiffMat}[1]{\ensuremath{\mat{\hat{\varPsi}}_{\Label{\delta},#1}}\xspace}

\newcommand{\SamplingFactor}{\ensuremath{N_\Label{S}}\xspace}

\newcommand{\RxTime}[3]{\ensuremath{t_{\Label{R},#1,#2}\!\left[#3\right]}\xspace}

\newcommand{\AnchorFrequency}[1]{\ensuremath{f_{\Label{a},#1}}\xspace}

\newcommand{\AlgoCurrentRxPulseIndex}[2]{\ensuremath{n_{\Label{R},#1,#2}}\xspace}

\newcommand{\AlgoLastTxPulseIndex}{\ensuremath{n_{\Label{T}}}\xspace}

\newcommand{\AlgoCurrentEstimatedRxTime}[2]{\ensuremath{\hat{t}_{\Label{R}, #1, #2}}\xspace}

\newcommand{\AlgoLastEstimatedRxTime}[2]{\ensuremath{\hat{t}_{\Label{LR}, #1, #2}}\xspace}

\newcommand{\AlgoLastTxTime}[1]{\ensuremath{t_{\Label{T},#1}}\xspace}

\newcommand{\AlgoCurrentRxTxTimeDiff}{\ensuremath{\tau_{\Label{\Delta}}}\xspace}

\newcommand{\AlgoCurrentRxTxPulseIndexDiff}{\ensuremath{n_{\Label{\Delta}}}\xspace}

\newcommand{\AlgoEstimatedPulseDuration}[2]{\ensuremath{\hat{T}_{#1,#2}}\xspace}

\newcommand{\AlgoAnchorDiffFrequency}[1]{\ensuremath{\varDelta_{\Label{a},#1}}\xspace}

\newcommand{\AlgoBaseFrequency}[1]{\ensuremath{f_{\Label{b},#1}}\xspace}

\newcommand{\AlgoAnchorFrequencyFactor}{\ensuremath{\alpha_\Label{a}}\xspace}

\newcommand{\CreateTerm}[3]{%
  \expandafter\newcommand\csname Term#1\endcsname{#2\xspace}%

  \expandafter\newcommand\csname Term#1P\endcsname{#3\xspace}%

  \expandafter\newcommand\csname Term#1C\endcsname{\makefirstuc{#2}\xspace}%

  \expandafter\newcommand\csname Term#1PC\endcsname{\makefirstuc{#3}\xspace}%
  
  \expandafter\newcommand\csname TermA#1\endcsname{%
    \IfBeginWith{#2}{a,e,i,o,u,A,E,I,O,U}{an #2\xspace}{a #2\xspace}%
  }%

  \expandafter\newcommand\csname TermA#1C\endcsname{%
    \IfBeginWith{#2}{a,e,i,o,u,A,E,I,O,U}{An #2\xspace}{A #2\xspace}%
  }%
}

\CreateTerm{BasePulse}{base signal}{base signals}
\CreateTerm{PulseDuration}{signal duration}{signal durations}
\CreateTerm{Undefined}{undefined}{undefineds}
\CreateTerm{BaseFrequency}{base frequency}{base frequencies}

\ifcsname func\endcsname
\else
    \newcommand{\func}[1]{\ensuremath{\mathrm{#1}}}
    \renewcommand{\vec}[1]{\ensuremath{\bm{#1}}}
    \newcommand{\mat}[1]{\vec{#1}}
\fi

\newcommand{\SimInstantaneousFrequency}[2]{\ensuremath{f_{#1}\!\left(#2\right)}\xspace}

\newcommand{\SimReferenceFrequency}{\ensuremath{f_{\Label{ref}}}\xspace}

\newcommand{\SimStartTimeError}[2]{\ensuremath{\func{err}_{\Label{start}, #1}\!\left( #2 \right)}\xspace}

\newcommand{\SimPositionEstimationError}[2]{\ensuremath{\func{err}_{\Label{pos}, #1}\!\left( #2 \right)}\xspace}

\newcommand{\SimTruePositionMatrix}{\ensuremath{\mat{X}}\xspace}

\newcommand{\SimEstimatePositionMatrix}[2]{\ensuremath{\mat{\hat{X}}_{p}\!\left( #2 \right)}\xspace}

\newcommand{\SimOptimalRotationMatrix}[2]{\ensuremath{\mat{C}\!\left( #1, #2\right)}\xspace}

\newcommand{\SimClosestTransmitIndex}[2]{\ensuremath{n_{\Label{closest},#1}\!\left(#2\right)}\xspace}

\newcommand{\SimulationFiguresPath}[1]{content/figures/simulation/#1/content}

\newcommand{\stepsection}[1]{%
  \vspace{0.5ex}%
  \noindent\textbf{#1}\par
  \vspace{0.5ex}%
}

\mathchardef\ordinarycolon\mathcode`\:
\mathcode`\:=\string"8000
\begingroup \catcode`\:=\active
  \gdef:{\mathrel{\mathop\ordinarycolon}}
\endgroup

\newcommand{\PaperTitle}{Synchronization and Localization in Ad-Hoc ICAS Networks Using a Two-Stage Kuramoto Method}
\newcommand{\PaperAcknlowledgement}{This work has been funded by the German Research Agency (DFG) under the project JCRS CoMP under project number 504990291.}
\newcommand{\PaperKeywords}{synchronization, syntonization, localization, ICAS, Kuramoto, frequency, phase}

\newcommand{\FirstAffiliationName}{Technische Universität Ilmenau}
\newcommand{\FirstAffiliationGroupName}{Radio Technologies for Automated and Connected Vehicles Research Group}
\newcommand{\FirstAffiliationPlace}{Ilmenau, Germany}

\newcommand{\FirstAuthorName}{Dominik Neudert-Schulz}
\newcommand{\FirstAuthorOrcid}{0009-0005-9230-3491}

\newcommand{\FirstAuthorAffiliationIndex}{1}

\newcommand{\SecondAuthorName}{Thomas Dallmann}
\newcommand{\SecondAuthorOrcid}{0000-0003-4655-6568}

\newcommand{\SecondAuthorAffiliationIndex}{1}

\begin{document}

\title{\PaperTitle\\
\thanks{\PaperAcknlowledgement}
}

\author{
    \IEEEauthorblockN{
		\ifedas
			\FirstAuthorName\IEEEauthorrefmark{\FirstAuthorAffiliationIndex},
			\SecondAuthorName\IEEEauthorrefmark{\SecondAuthorAffiliationIndex}
		\else
			\orcidlinki{\FirstAuthorName\IEEEauthorrefmark{\FirstAuthorAffiliationIndex}}{\FirstAuthorOrcid},
			\orcidlinki{\SecondAuthorName\IEEEauthorrefmark{\SecondAuthorAffiliationIndex}}{\SecondAuthorOrcid}
		\fi
    }
    \IEEEauthorblockA{
        \{first name\}.\{last name\}@tu-ilmenau.de\\
        \IEEEauthorrefmark{1}\textit{\FirstAffiliationGroupName}\\
        \textit{\FirstAffiliationName}\\
        \FirstAffiliationPlace{}
    }
}

\maketitle

\begin{abstract}
To enable Integrated Communications and Sensing (ICAS) in a peer-to-peer vehicular network,
precise synchronization in frequency and phase among the communicating entities is required.
In addition, self-driving cars need accurate position estimates of the surrounding vehicles.
In this work, we propose a joint, distributed synchronization and localization scheme for a network of communicating entities.
Our proposed scheme is mostly signal-agnostic and therefore can be applied to a wide range of possible ICAS signals.
We also mitigate the effect of finite sampling frequencies,
which otherwise would degrade the synchronization and localization performance severely. 

\end{abstract}


\section{Introduction}

Vehicular networks are envisioned to be able to support Integrated Communications and Sensing (ICAS).
That is, data transmission and radar applications shall be possible using the same signal.
Such signals may be communication signals like OFDM (communication-centric ICAS)
or dedicated radar signals like FMCW or pulse radar (radar-centric ICAS).
Hence, there is a rich zoo of signals that might be considered for future ICAS applications.

In situations where there is no central coordinating unit, the
communicating entities, from now on called \emph{hosts}, need to cooperate on a peer-to-peer (P2P) basis.
In order to allow for the most accurate radar applications, coherent signal processing is required.~\cite{Elga25}
For P2P scenarios, this means that hosts need to be synchronized exactly in both, frequency and phase.
Moreover, precise position information may be required to support self-driving applications.
However, especially in urban scenarios low-cost global navigation satellite systems (GNSS) receivers
often fail to provide exact position estimates.\cite{GroAdj19,AdjGroQuickEll19}

Joint phase and frequency synchronization is a long-standing problem.
Consensus based algorithms as proposed in~\cite{Dall21},~\cite{Ba23}, and~\cite{BaHeDall24}
are able to solve this problem in a fully distributed manner. 
However, they either neglect the propagation delay completely or do not contribute to position estimation.

This work builds on the synchronization schemes proposed in~\cite{BaHeDall24} and~\cite{Dall21}.
The basis is the Kuramoto model~\cite{Ku75},
that is used to describe the convergence behavior of a large number of coupled oscillators.
In~\cite{Dall21}, it has been shown that it is possible to obtain frequency consensus in a network, 
that is designed to behave as a network of coupled oscillators. 
While the phase relation become stable, the phases themselves will not converge to a consensus. 
This problem is solved by the two-stage Kuramoto model described in~\cite{BaHeDall24}.
However, both works ignore the effect of finite sampling frequencies and propagation delays in real-world applications.
We close this gap making the two-stage Kuramoto method applicable to real-world scenarios.
Since our proposed method also estimates the propagation delays between the hosts,
we are able to perform mutual localization of all hosts.

This work is organized as follows:
In \abschnitt{system_model}, we describe the signal model and a generalized phase definition.
In \abschnitt{synchronization}, we review the two-stage Kuramoto method and show how to acquire the required phase differences.
\Abschnitt{algorithm} outlines our proposed synchronization and localization algorithm.
In \abschnitt{simulation}, we use simulations to validate our algorithm in different scenarios.
Finally, \abschnitt{conclusion} concludes this work.

Notation: 
Throughout this work, vectors and matrices are denoted by bold lower-case and upper-case letters, respectively.
A hat~$\hat{}$ on top of a variable denotes the estimate of the corresponding variable.
We will use ${}\mTr$ and \mFrobNorm{\cdot} to denote the matrix transpose and the
Frobenius norm, respectively.

\section{System model}\label{sec:system_model}
\begin{figure}[t!]
    \centering
    \includegraphics[width=\columnwidth]{content/figures/signal_sketch/figure.tikz}
    \caption{Exemplary transmit signal \TxSignal{p}{t} of some host $p$ consisting of several \TermBasePulseP \BasePulse{p}{n}{t}}
    \label{fig:signalsketch}
\end{figure}

In this section, we will highlight the signal model and
derive the phase difference definition used in the following sections.

\subsection{Signal model}

We consider a number of $P$ simultaneously transmitting and receiving hosts.
Each host transmits a signal 
$
\TxSignal{p}{t}: \setR \to \setC
$, 
where $p$ is the index of the host. 
This signal is a concatenation of \emph{\TermBasePulseP{}}
$
\BasePulse{p}{n}{u}: \setR \to \setC
$, with $n \in \setN_0$.
The actual form of the \TermBasePulseP{} \BasePulse{p}{n}{u} is not important 
as long as the following constraints are guaranteed:
\begin{enumerate}
    \item A receiving host must be able to deduce the host which transmitted \TermABasePulse{}.
    \item A receiving host must be able to determine the reception time of each transmitted \TermBasePulse{} \BasePulse{p}{n}{u} with sufficient precision.
          This must hold true even if multiple hosts are transmitting simultaneously and even if the \TermBasePulseP{} overlap in time due to multipath propagation.
    \item The \TermBasePulseP{} are causal ($\BasePulse{p}{n}{u} = 0$ for $u < 0$)
          and have unit duration ($\BasePulse{p}{n}{u} = 0$ for $u \geq 1$).
\end{enumerate}

These constraints are often already fulfilled by practically used signals.
Hence, the synchronization and localization algorithm presented later on is applicable to existing systems.



Let $T_{p, n}$ be the duration of the $n$-th \TermBasePulse{} of host $p$ 
and let \FirstPulseStartTime{p} be the time of the first \TermBasePulse{} transmitted by host $p$. 
The overall signal transmitted by host $p$ is a concatenation of stretched or compressed \TermBasePulseP{}.
\begin{align}
    \TxSignal{p}{t} &:= 
        \sum_{n=0}^\infty \BasePulse{p}{n}{
            \frac{t - \PulseStartTime{p}{n}}{\PulseDuration{p}{n}} 
        }
    \\
    \PulseStartTime{p}{n} &:= \PulseStartTime{p}{n-1} + \PulseDuration{p}{n-1} = \FirstPulseStartTime{p} + \sum_{k=1}^{n-1} \PulseDuration{p}{k}
    \\
    \PulseStartTime{p}{0} &:= \FirstPulseStartTime{p}, \nonumber
    \qquad
    \PulseDuration{p}{0} := \FirstPulseDuration{p} \nonumber
\end{align}
An exemplary depiction of such a signal \TxSignal{p}{t} for some host $p$ is shown in \abbildung{signalsketch}.

Even though a host may send a series of different \TermBasePulseP{}, 
we regard this model as a kind of quasi-periodic function where the \TermPulseDuration{} \PulseDuration{p}{n} may change with every \TermBasePulse{}.

On the receiving side, we consider a signal model, where some host $q$ receives $K_{p,q}$ copies of the signal \TxSignal{p}{t} transmitted by host $p$.
These copies are attenuated by $\alpha_{p,q,k_p}$ and delayed by \PropagationDuration{p}{q}{k_p}{t}, where $k_p = 0 \ldots K_{p,q} - 1$.
\begin{align}
    \func r_q(t) 
        := 
            \sum_{\substack{p=0\\p \neq q}}^{P-1} 
            \sum_{k_p=0}^{K_{p,q}} 
            \alpha_{p,q,k_p}\, \func x_p(t - \PropagationDuration{p}{q}{k_p}{t})
\end{align}
In this model, $k_p=0$ corresponds to the line of sight (LOS) component.
We presume that $\alpha_{p,q,0}$ is always sufficiently large such that 
for any pair of receiving host $q$ and transmitting host $p$ ($p \neq q$)
the LOS component of the $n_{p, q}$-th \TermBasePulse{} from host $p$ to host $q$ 
can be detected and its reception time 
$
    \RxTime{p}{q}{n_{p, q}} := \PulseStartTime{p}{n_{p, q}} + \PropagationDuration{p}{q}{0}{t}
$
can be estimated.

As stated above, all hosts transmit their sequences of \TermBasePulseP{}.
Every host starts to transmit at a different point in time.
Moreover, each host has its own clock which is usually not synchronized in time with the other hosts.
Our goal is to synchronize all hosts ($p = 0 \ldots P-1$) so that their \TermPulseDurationP{} \PulseDuration{p}{n_p}
and their \TermBasePulse{} start times \PulseStartTime{p}{n_p} each converge to a common value across all hosts
as $n_p\to\infty$. 
Additionally, we will use the synchronization process to estimate the relative positions of hosts.

\subsection{Phases and phase differences}

We aim to synchronize the waveforms in \TermPulseDuration{} and in pulse start times by
utilizing the two-stage Kuramoto method.
This method relies on a notion of a \emph{phase}, for which we will now provide a definition tailored to our signal model.

A classical approach to define a phase involves the construction of an analytical signal 
$\func x_{\Label{analytic},p} := \func x_p(t) + \jmath \mathcal{H} \left\lbrace \func x_p(t) \right\rbrace$ 
using the Hilbert transform $\mathcal H$. 
In this case, the phase would be the angle in the complex plane $\varphi_{\Label{analytic},p}(t) = \arg\left( \func x_{\Label{analytic},p}(t) \right)$
However, this only works well for narrowband signals. Wideband signals could exhibit sudden spikes in $\varphi_{\Label{analytic},p}(t)$.
Therefore, we resort to an alternative phase definition.

Let \TxPulseIndex{q}{t} be the index of the \TermBasePulse{} being transmitted by host $q$ at time instant $t$.
We define the \emph{transmit phase} of $\func x_q(t)$ as follows.
\begin{align}\label{eq:TransmitPhase}
    \TxPhase{q}{t} := 
        2 \pi\!\left(
            \TxPulseIndex{q}{t} 
            + 
            \frac{ 
                t - \PulseStartTime{q}{\TxPulseIndex{q}{t}} 
            }{ 
                \PulseDuration{q}{\TxPulseIndex{q}{t}}
            } 
        \right)
\end{align}

The same idea is applied to the LOS signal $\func x_p(t - \PropagationDuration{p}{q}{0}{t})$ from host $p$ to host $q$.
At some time instant $t$, let \RxPulseIndex{p}{q}{t} be the index of the last \TermBasePulse{} that host $q$ received from host $p$,
Then, we define the \emph{receive phase} of $\func x_p(t - \delta_{p,q,0}(t))$ as:
\begin{align}\label{eq:ReceivePhase}
    \RxPhase{p}{q}{t} 
        :=  
            2 \pi\!\left(
                \RxPulseIndex{p}{q}{t} 
                + 
                \frac{
                    t - \PropagationDuration{p}{q}{0}{t} - \PulseStartTime{p}{\RxPulseIndex{p}{q}{t}}
                }{
                    \PulseDuration{p}{\RxPulseIndex{p}{q}{t}}
                } 
            \right)
\end{align}

The phase definitions presented in \gleichung{TransmitPhase} and \gleichung{ReceivePhase} are designed to
have two main properties:
First, the phase increases linearly from $0$ to $2 \pi$ within each \TermBasePulse{}.
Hence, the lengths of previous \TermBasePulseP{} do not influence the phase {-} only their number does.
Second, besides the factor of $2\pi$ the slope of the phase within each \TermBasePulse{} is an instantaneous frequency 
(i.e., inverse of \TermPulseDuration{}).



The synchronization and localization approach presented in this work
relies on the knowledge of a phase difference between the signal $\func x_q(t)$ sent by host $q$ 
and the LOS signal $\func x_p(t - \PropagationDuration{p}{q}{0}{t})$ sent by host $p$ and received at host $q$.
Note that pure synchronization would be possible with non-LOS components as well.
Even a combination of multiple components may be possible.
However, only the LOS allows us to estimate the distances between each host 
and therefore render localization possible.

Formally, the phase difference \RealRxPhaseDiff{p}{q}{t} between hosts $q$ and $p$ at time instant $t$ could be defined as
$
    \RealRxPhaseDiff{p}{q}{t} :=
        \TxPhase{q}{t}
        -
        \RxPhase{p}{q}{t}.
$
Unfortunately, \RxPhase{p}{q}{t} includes the \TermPulseDuration{} \PulseDuration{p}{\RxPulseIndex{p}{q}{t}} of host $p$,
which is not known at host $q$.
One solution would be to use the previous \TermPulseDuration{} of host $p$ estimated by host $q$. 
Another approach presented in \cite{Dall21} simplifies the phase difference by assuming that host $q$'s \TermPulseDuration{} 
is roughly equal to host $p$'s \TermPulseDuration{} leading to the approximation
\begin{align}
    \RealRxPhaseDiff{p}{q}{t} \approx\,
        2 \pi 
        \left(
            \RxTxPulseIndexDiff{p}{q}{t}
            + 
            \frac{
                \RxTxPulseTimeDiff{p}{q}{t}
                +
                \PropagationDuration{p}{q}{0}{t}
            }{
                \PulseDuration{q}{\TxPulseIndex{q}{t}}
            } 
        \right),
\end{align}
where
\begin{align*}
    \RxTxPulseTimeDiff{p}{q}{t}
        :=\,& 
            \PulseStartTime{p}{\RxPulseIndex{p}{q}{t}}
            - 
            \PulseStartTime{q}{\TxPulseIndex{q}{t}}    
    \\
    \RxTxPulseIndexDiff{p}{q}{t}
        :=\,&
            \TxPulseIndex{q}{t} 
            - 
            \RxPulseIndex{p}{q}{t}.
\end{align*}

In order to improve the accuracy of the phase difference estimate, a sampling factor \SamplingFactor is introduced in \cite{Dall21},
that includes multiple \TermBasePulseP{} into a single effective phase difference \RxPhaseDiff{p}{q}{t}.
\begin{align}\label{eq:FinalPhaseDiff}
    \RxPhaseDiff{p}{q}{t} :=\,
        \frac{2 \pi}{\SamplingFactor}
        \left(
            \RxTxPulseIndexDiff{p}{q}{t}
            + 
            \frac{
                \RxTxPulseTimeDiff{p}{q}{t}
                +
                \PropagationDuration{p}{q}{0}{t}
            }{
                \PulseDuration{q}{\TxPulseIndex{q}{t}}
            } 
        \right)
\end{align}

\section{Synchronization and localization}\label{sec:synchronization}

In this section, we review the two-stage Kuramoto method.
After that, we show how we can estimate propagation delays 
based on the phase differences defined in \gleichung{FinalPhaseDiff}.

\subsection{The two-stage Kuramoto method}

The two-stage Kuramoto method proposed in~\cite{BaHeDall24}
can be used to synchronize hosts in frequency and phase.
This method assumes that each host transmits at an instantaneous frequency $f_q(t)$, with $q = 0 \ldots P-1$,
that can be decomposed into a base frequency $f_{\Label{b}, q}(t)$ and 
into an update term that depends on a phase differences $\varDelta_{q, n_q}(t)$.

The first stage consists of a simple consensus filter which updates the base frequency.
\begin{align}
    \deriv{f_{\Label{b}, q}(t)}{t} = -\epsilon \sum_{\substack{p=0\\p \neq q}}^{N-1}  \left( f_{\Label{b}, q}(t) - f_{\Label{b}, p}(t) \right)
\end{align}
The second stage describes the evolution of the instantaneous frequency $f_q(t)$. 
It resembles a first-order Kuramoto model with the intrinsic natural frequency being replaced by the base frequency $f_{\Label{b}, q}(t) $.
\begin{align}
    f_q(t) 
        = 
            f_{\Label{b}, q}(t) 
            - \varepsilon \sum_{\substack{p=0\\p \neq q}}^{P-1} 
                \sin\left(
                    \varDelta_{q, n_q}(t)
                \right)
\end{align}

The authors of~\cite{BaHeDall24} could show that their two-stage Kuramoto method is capable of synchronizing the frequencies and the phases of a number of entities in a network.
In the following sections, we will extend this method to add localization capabilities.
Moreover, we will extend the first stage such that the method works also in the presence of sampling effects.

\subsection{Estimating transmit times and propagation delays}

The phase difference \RxPhaseDiff{p}{q}{t} defined in \gleichung{FinalPhaseDiff} includes the LOS propagation delay \PropagationDuration{p}{q}{0}{t}.
Hence, we cannot use \RxPhaseDiff{p}{q}{t} directly for frequency and phase synchronization.

To combat this, we decompose \RxPhaseDiff{p}{q}{t} into a phase difference \PulseStartTimeDepRxPhaseDiff{p}{q}{t} that only
depends on the \TermBasePulse{} start times and into a phase difference \PropagationTimeDepRxPhaseDiff{p}{q}{t}
that includes the propagation delay:
\begin{align}\label{eq:phase_diff_sum}
    \RxPhaseDiff{p}{q}{t}
        :=
            \PulseStartTimeDepRxPhaseDiff{p}{q}{t}
            +
            \PropagationTimeDepRxPhaseDiff{p}{q}{t},
\end{align}
where
\begin{align}
    \PulseStartTimeDepRxPhaseDiff{p}{q}{t}
        :=\,&
           \frac{2 \pi}{\SamplingFactor}
            \left(
                \RxTxPulseIndexDiff{p}{q}{t}
                + 
                \frac{
                    \RxTxPulseTimeDiff{p}{q}{t}
                }{
                    \PulseDuration{q}{\TxPulseIndex{q}{t}}
                } 
            \right) 
    \\
    \PropagationTimeDepRxPhaseDiff{p}{q}{t}
        :=\,&
            2 \pi
            \frac{    
                \PropagationDuration{p}{q}{0}{t}
            }{
                \SamplingFactor    
                \PulseDuration{q}{\TxPulseIndex{q}{t}}                    
            } 
\end{align}

The joint synchronization and localization algorithm described in the following section
requires knowledge of
\PulseStartTimeDepRxPhaseDiff{p}{q}{t} and \PropagationTimeDepRxPhaseDiff{p}{q}{t}.
Hence, we will now analyze, how the decomposition given in \gleichung{phase_diff_sum} can be computed.
To this end, we rewrite \gleichung{phase_diff_sum} in terms of matrices as
\begin{align}\label{eq:phase_diff_matrix_sum}   
    \RxPhaseDiffMat
        =\  
            \PulseStartTimeDepRxPhaseDiffMat
            +
            \PropagationTimeDepRxPhaseDiffMat,
\end{align}
where
\begin{align*}
    \left[ \RxPhaseDiffMat(t) \right]_{p,q} :=\,& \RxPhaseDiff{p}{q}{t},
    &
    \left[ \PulseStartTimeDepRxPhaseDiffMat(t) \right]_{p,q} :=\,& \PulseStartTimeDepRxPhaseDiff{p}{q}{t},
    \\
    \left[\PropagationTimeDepRxPhaseDiffMat(t)\right]_{p,q} :=\,& \PropagationTimeDepRxPhaseDiff{p}{q}{t}.
\end{align*}
For notational convenience, we will omit writing the dependency of these matrices on the time $t$.

Suppose that no \TermBasePulse{} is lost during transmission. 
That is, all \TermBasePulseP{} transmitted by one host are eventually received by all other hosts.
In this case, the number of transmitted \TermBasePulseP{} \TxPulseIndex{p}{t} equals the number of received \TermBasePulseP{} \RxPulseIndex{p}{q}{t}.
It follows that $\RxTxPulseIndexDiff{p}{q}{t} = - \RxTxPulseIndexDiff{q}{p}{t}$.
In the limit of perfectly synchronized pulse start times ($\RxTxPulseTimeDiff{p}{q}{t} = 0$),
it holds that $\PulseStartTimeDepRxPhaseDiff{p}{q}{t} = - \PulseStartTimeDepRxPhaseDiff{q}{p}{t}$.
Hence, \PulseStartTimeDepRxPhaseDiffMat becomes a skew-symmetric matrix 
($
\PulseStartTimeDepRxPhaseDiffMat = - \PulseStartTimeDepRxPhaseDiffMat\mTr
$).

Next, we exploit the fact that propagation delays are the same in both directions 
($
\PropagationDuration{p}{q}{0}{t}\!=\!\PropagationDuration{q}{p}{0}{t}
$).
In the limit of perfectly synchronized pulse durations
($
\PulseDuration{p}{\TxPulseIndex{p}{t}}\!=\!\PulseDuration{q}{\TxPulseIndex{q}{t}}
$)
the phase differences \PropagationTimeDepRxPhaseDiff{p}{q}{t} and \PropagationTimeDepRxPhaseDiff{q}{p}{t} become equal.
Hence, \PropagationTimeDepRxPhaseDiffMat{} becomes a symmetric matrix
($
\PropagationTimeDepRxPhaseDiffMat = \PropagationTimeDepRxPhaseDiffMat\mTr
$). 

Finally, consider the case where both, the \TermBasePulse{} start times and \TermPulseDurationP{}, are sufficiently synchronized,
such that 
$
\PulseStartTimeDepRxPhaseDiffMat \approx - \PulseStartTimeDepRxPhaseDiffMat\mTr
$
and
$
\PropagationTimeDepRxPhaseDiffMat \approx \PropagationTimeDepRxPhaseDiffMat\mTr
$.
In this case, the decomposition given in \gleichung{phase_diff_matrix_sum} can be estimated as follows.
\begin{align}\label{eq:phase_matrices}
    \PulseStartTimeDepRxPhaseDiffMat
        \approx& 
            \tfrac{1}{2}
            \left(
                \RxPhaseDiffMat - \RxPhaseDiffMat\mTr
            \right), 
    &            
    \PropagationTimeDepRxPhaseDiffMat
        \approx& 
            \tfrac{1}{2}
            \left(
                \RxPhaseDiffMat + \RxPhaseDiffMat\mTr
            \right)
\end{align}

\section{Proposed algorithm}\label{sec:algorithm}

\begin{figure}[t]
    \centering
    \includegraphics[width=\columnwidth]{content/figures/algorithm_variables/figure.tikz}
    \caption{Exemplary scenario as seen from host $q$: Host $q$ receives \TermBasePulseP{} from host $p$ and transmits its own \TermBasePulseP{}.
    Vertical lines indicate the point in time of transmission (red/circle) and reception (blue/square), respectively.}
    \label{fig:algorithm_variables}
\end{figure}

In this section, we will explain our proposed synchronization process,
that is executed on each host in order to continuously adapt the \TermPulseDuration{}.
To this end, we assume a fully connected network, 
where all hosts listen to the \TermBasePulseP{} transmitted by all other hosts.

\subsection{Notation}

In order to avoid an unnecessarily complex description of the proposed algorithm,
we will resort to a simplified notation for the following subsections.
Especially, we will omit the time dependency of various parameters,
assuming the current point in time of reception or transmission, respectively.
Moreover, we will use the following quantities:
Let \TxPulseIndex{q}{} denote the index of the \TermBasePulse currently being transmitted by host $q$.
This \TermBasePulse{} started at time \AlgoLastTxTime{q} and has duration \PulseDuration{q}{\TxPulseIndex{q}{}}. 
Moreover, let \AlgoCurrentEstimatedRxTime{p}{q} and \AlgoLastEstimatedRxTime{p}{q}
denote the time 
-- measured in host $q$'s time frame --
at which the current \TermBasePulse and the last \TermBasePulse from host $p$ arrived at host $q$, respectively.
Host $q$'s index of the current \TermBasePulse{} received from host $p$  is \AlgoCurrentRxPulseIndex{p}{q}.
\Abbildung{algorithm_variables} shows an exemplary scenario.

\subsection{Distribution of estimated phase differences}

During the synchronization process, each host $p$ estimates a vector \EstimatedRxPhaseDiffVec{p} of phase differences
using \gleichung{FinalPhaseDiff}.
This vector is transmitted to all other hosts either in-band (modulated on the \TermBasePulseP{}) or via a separate communication channel (out-of-band).
Note that, in order to save network resources, this vector may be transmitted less frequently than  the \TermBasePulseP{}.

When a host $q$ receives such a vector from host $p$, host $q$ updates the $p$-th column
of its estimated phase difference matrix $\EstimatedRxPhaseDiffMat{q}$ 
(host $q$'s estimate of \RxPhaseDiffMat).
\begin{align}
    \left[ \EstimatedRxPhaseDiffMat{q} \right]_{:,p} 
        \leftarrow\,
        \EstimatedRxPhaseDiffVec{p}
\end{align}

The $q$-th column of \EstimatedRxPhaseDiffMat{q} is updated upon reception of \TermBasePulseP{}
a described in the next subsection.

\subsection{\TermBasePulseC{} reception}

Upon reception of \TermABasePulse{} from some host $p$ at time instant $\AlgoCurrentEstimatedRxTime{p}{q}$,
host $q$ estimates the phase difference between itself and host $p$. 
The estimated phase difference is stored in the matrix \EstimatedRxPhaseDiffMat{q}.
Based on \gleichung{FinalPhaseDiff}, the phase difference is estimated as
\begin{align}
        \left[ \EstimatedRxPhaseDiffMat{q} \right]_{p,q}
    \leftarrow\,
        \frac{2 \pi}{\SamplingFactor}
        \left(
            \AlgoCurrentRxTxPulseIndexDiff
            + 
            \frac{
                \AlgoCurrentRxTxTimeDiff
            }{
                \PulseDuration{q}{\AlgoLastTxPulseIndex}
            } 
        \right),
\end{align}
where 
$
\AlgoCurrentRxTxTimeDiff 
    := \AlgoCurrentEstimatedRxTime{p}{q} - \AlgoLastTxTime{q}
$
and
$
\AlgoCurrentRxTxPulseIndexDiff 
    := \TxPulseIndex{q}{} - \AlgoCurrentRxPulseIndex{p}{q}
$.

In order to conduct the synchronization, host $q$ needs to estimate the \TermPulseDurationP{} of all other hosts.
To this end, every time host $q$ receives \TermABasePulse{} from host $p$,
it updates its estimate of host $p$'s \TermPulseDuration{} \AlgoEstimatedPulseDuration{p}{q}
using the current reception time \AlgoCurrentEstimatedRxTime{p}{q} and the previous reception time \AlgoLastEstimatedRxTime{p}{q}.
\begin{align}
    \AlgoEstimatedPulseDuration{p}{q}
        \leftarrow\,
            \AlgoCurrentEstimatedRxTime{p}{q} 
            - 
            \AlgoLastEstimatedRxTime{p}{q}
\end{align}

\subsection{Localization}
Once all columns of host $q$'s phase difference matrix \EstimatedRxPhaseDiffMat{q} are known, 
host $q$ can estimate the phase matrix \EstimatedPropagationTimeDepRxPhaseDiffMat{q}
according to \gleichung{phase_matrices}. After this, host $q$ can utilize the
propagation delays in \EstimatedPropagationTimeDepRxPhaseDiffMat{q} to estimate relative
positions of all other hosts using techniques like multidimensional scaling~\cite{Rich38,Tor52}.

\subsection{Synchronization}
The synchronization step estimates the duration \PulseDuration{q}{\TxPulseIndex{q}{}+1} of 
the next \TermBasePulse{} to be transmitted by host $q$.
This step requires that host $q$ has estimated the \TermPulseDurationP \AlgoEstimatedPulseDuration{p}{q}
of all other hosts $p$ as well as the $q$-th column of \EstimatedPulseStartTimeDepRxPhaseDiffMat{q}.
The matrix \EstimatedPulseStartTimeDepRxPhaseDiffMat{q} may either be computed 
via \gleichung{phase_matrices} or by subtracting \EstimatedPropagationTimeDepRxPhaseDiffMat{q}
from \EstimatedRxPhaseDiffMat{q}.

\stepsection{First stage of Kuramoto method}

This sub-step corresponds to the first stage of the two-stage Kuramoto method presented in~\cite{BaHeDall24} 
with an additional penalty term to prevent drift of the \TermPulseDurationP{}.

In this step, host $q$ computes \TermABaseFrequency \AlgoBaseFrequency{q}
using its estimated \TermPulseDurationP{} \AlgoEstimatedPulseDuration{p}{q},
where \AlgoEstimatedPulseDuration{q}{q} shall be set to host $q$'s own current \TermPulseDuration{}.
\begin{align}\label{eq:BaseFrequency}
    \AlgoBaseFrequency{q}
        \leftarrow\, 
            \frac{1}{P} \sum_{p=0}^{P-1} {\AlgoEstimatedPulseDuration{p}{q}}^{-1}
            -
            \AlgoAnchorFrequencyFactor{} \AlgoAnchorDiffFrequency{q}{t}
\end{align}
The quantity \AlgoAnchorDiffFrequency{q} is a penalty that comes into play if the \TermPulseDurationP of all
hosts jointly drift away. For this, host $q$ has to choose an anchor frequency \AnchorFrequency{q}.
\begin{align}
    \AlgoAnchorDiffFrequency{q}
        \leftarrow\,
            \frac{1}{P} \sum_{p=0}^{P-1} {\AlgoEstimatedPulseDuration{p}{q}}^{-1}
            -
            \AnchorFrequency{q}
\end{align}
The anchor frequency \AnchorFrequency{q} is assumed to be constant over the synchronization period.
In this work, we chose \AnchorFrequency{q} to be the average over the initial \TermPulseDurationP{}
\EstimatedFirstPulseDuration{p}{q}, 
where \EstimatedFirstPulseDuration{p}{q} is host $q$'s estimate of the true initial pulse duration \FirstPulseDuration{p}.
\begin{align}
    \AnchorFrequency{q}
        :=\,
            \frac{1}{P} \sum_{p=0}^{P-1} {\EstimatedFirstPulseDuration{p}{q}}^{-1} 
\end{align}
The factor \AlgoAnchorFrequencyFactor{} should be a small positive value close to zero.

\stepsection{Second stage of Kuramoto method}

The first stage of the Kuramoto method achieves synchronization of the \TermPulseDurationP \PulseDuration{q}{n}.
In order to synchronize the \TermBasePulse{} start times \PulseStartTime{q}{n} we use the second stage of the two-stage Kuramoto method,
which defines the \TermPulseDuration{} \PulseDuration{q}{\TxPulseIndex{q}{}+1} used for the next \TermBasePulse{}.
\begin{align}\label{eq:FinalPulseDuration}
    \PulseDuration{q}{\TxPulseIndex{q}{}+1}
        \leftarrow\,
            \left(
                \AlgoBaseFrequency{q}
                - 
                \frac{K}{P} \sum_{p=0}^{P-1} \sin\!\left( \left[ \EstimatedPulseStartTimeDepRxPhaseDiffMat{q} \right]_{p,q} \right)
            \right)\mInv
\end{align}
Here, the main diagonal of \EstimatedPulseStartTimeDepRxPhaseDiffMat{q} is defined to be zero.

\section{Simulation}\label{sec:simulation}


\newcommand{\SimValueSamplingFrequency}{800.00~MHz}

\newcommand{\SimValueSamplingFactor}{8}

\newcommand{\SimValueHostCount}{4}

\newcommand{\SimValueHostZeroPos}{\ensuremath{\left[0~\text{m}, 0~\text{m}, 0~\text{m}\right]\mTr}}

\newcommand{\SimValueHostOnePos}{\ensuremath{\left[30~\text{m}, 10~\text{m}, 0~\text{m}\right]\mTr}}

\newcommand{\SimValueHostTwoPos}{\ensuremath{\left[10~\text{m}, -30~\text{m}, -10~\text{m}\right]\mTr}}

\newcommand{\SimValueHostThreePos}{\ensuremath{\left[20~\text{m}, -25~\text{m}, 20~\text{m}\right]\mTr}}

In this section, we will validate our proposed synchronization and localization scheme by means of simulations.

\subsection{Simulation setup}

We simulated four hosts.
Their initial \TermPulseDuration, initial transmit start times, and their positions are listed in Table~\ref{tab:hostsimparams}. 
Each host transmits a \TermBasePulse and performs the proposed algorithm.
The simulation assumes ideal conditions without noise or interference.


\begin{table}[H]
  \centering
  \begin{tabular}{llll}
    \toprule
    host&\FirstPulseDuration{\cdot} / \FirstPulseFrequency{\cdot}&\FirstPulseStartTime{\cdot}&position\\ 
    \midrule
0 & 142.86~ns / 7.00~MHz & 64.29~ns & \ensuremath{\left[0~\text{m}, 0~\text{m}, 0~\text{m}\right]\mTr} \\ 
1 & 125.00~ns / 8.00~MHz & 6.25~ns & \ensuremath{\left[30~\text{m}, 10~\text{m}, 0~\text{m}\right]\mTr} \\ 
2 & 136.99~ns / 7.30~MHz & 22.60~ns & \ensuremath{\left[10~\text{m}, -30~\text{m}, -10~\text{m}\right]\mTr} \\ 
3 & 131.58~ns / 7.60~MHz & 32.89~ns & \ensuremath{\left[20~\text{m}, -25~\text{m}, 20~\text{m}\right]\mTr} \\ 
    \bottomrule
  \end{tabular}
  \caption{Host parameters}\label{tab:hostsimparams}
\end{table}

To show the effect of the sampling on the proposed synchronization approach, we conducted simulations with three different scenarios.
In scenario A, we assumed perfect sampling without any error (i.e., infinite sampling frequency).
In scenario B, we assumed a finite sampling frequency but no drift compensation ($\AlgoAnchorFrequencyFactor = 0$).
In scenario C, we also assumed a finite sampling frequency but this time with applied drift compensation.
\Tabelle{simscenarios} summarizes the different scenarios and their parameters.

\begin{table}[H]
  \centering
  \begin{tabular}{llll}
    \toprule
    scenario & \SamplingFrequency & \AlgoAnchorFrequencyFactor{} & comment\\ 
    \midrule
      A & $\infty$ Hz & 0 & no sampling error\\
      B & \SimValueSamplingFrequency{} & 0 & no drift compensation\\
      C & \SimValueSamplingFrequency{} & 0.05 & with drift compensation\\
    \bottomrule
  \end{tabular}
  \caption{Simulated scenarios}\label{tab:simscenarios}
\end{table}

\subsection{Performance metrics}

For each simulation setup, we will analyze the synchronization in phase and frequency as well as the localization.

The first plot (a) of each scenario depicts the evolution of the instantaneous frequency 
$
\SimInstantaneousFrequency{p}{t} 
    := 
        1 / \PulseDuration{p}{\TxPulseIndex{p}{t}}
$
of each host $p$ over time.
The left axis shows the actual instantaneous frequency, 
while the right axis shows the difference to a reference frequency \SimReferenceFrequency.
The reference frequency is provided below each plot (a).

The phase synchronization is measured via the transmit start times \PulseStartTime{p}{\TxPulseIndex{p}{t}} of each host.
In plot (b) of each scenario, we depict the error function \SimStartTimeError{p}{t}.
\begin{align}    
    \SimStartTimeError{p}{t} 
        :=&
            \frac{1}{P-1}        
            \sum_{\stackrel{q=0}{q \neq p}}^{P-1}
                \left|
                    \PulseStartTime{p}{\TxPulseIndex{p}{t}}
                    -
                    \PulseStartTime{q}{\SimClosestTransmitIndex{q}{t}}
                \right|
    \\
    \SimClosestTransmitIndex{q}{t} 
        :=&         
            \argmin_{n}\!\left|
                    t
                    -
                    \PulseStartTime{q}{n}
                \right|
\end{align}
For each transmit time \PulseStartTime{p}{\TxPulseIndex{p}{t}} of some host $p$, \SimStartTimeError{p}{t} looks for the nearest transmit time of all other hosts and computes the mean absolute difference.
This value should go to zero as the hosts synchronize their transmit times.

Finally, we analyze the localization error by looking at the relative position error \SimPositionEstimationError{p}{t}.
\begin{align}
    \SimPositionEstimationError{p}{t} 
        := \frac{
            \mFrobNorm{
                \SimTruePositionMatrix
                -
                \SimOptimalRotationMatrix{\SimTruePositionMatrix}{\SimEstimatePositionMatrix{p}{t}} 
                \SimEstimatePositionMatrix{p}{t}
            } 
        }{
            \mFrobNorm{\SimTruePositionMatrix}
        }
\end{align}
Here, \SimTruePositionMatrix{} is a matrix whose columns contain the true host positions and 
\SimEstimatePositionMatrix{p}{t} is host $p$'s estimate of the position matrix at time $t$.
The estimated position matrices \SimEstimatePositionMatrix{p}{t} contain only relative positions.
Therefore, we pre-multiply \SimEstimatePositionMatrix{p}{t} by a rotation matrix \SimOptimalRotationMatrix{\mat A}{\mat B}
computed via Procrustes analysis such that
$
    \mFrobNorm{
        \mat A
        -
        \SimOptimalRotationMatrix{\mat A}{\mat B}
        \mat B
    }
$
is minimized (cf.~\cite{Schoe66}).

    \begingroup

    \newcommand{\PlotReferenceFrequencyValue}{6662.443\xspace}

\newcommand{\PlotReferenceFrequencyUnit}{kHz\xspace}



    \begin{figure}[t]%
        \centering
        \begin{subfigure}{\columnwidth}
            \resizebox{\columnwidth}{!}{\input{\SimulationFiguresPath{}/instantaneous_frequencies.pgf}}%
            \caption{Instantaneous frequency \SimInstantaneousFrequency{p}{t} ($\SimReferenceFrequency = \PlotReferenceFrequencyValue\;\text{\PlotReferenceFrequencyUnit}$)}\label{fig:C_instfreq}
        \end{subfigure}

        \begin{subfigure}{\columnwidth}
            \resizebox{\columnwidth}{!}{\input{\SimulationFiguresPath{}/transmit_deviations.pgf}}
            \caption{Mean absolute differences of \TermBasePulse{} start times}\label{fig:C_txdev}
        \end{subfigure}

        \begin{subfigure}{\columnwidth}
            \resizebox{\columnwidth}{!}{\input{\SimulationFiguresPath{}/relative_position_errors.pgf}}
            \caption{Relative position error}\label{fig:C_poserr}
        \end{subfigure}

        \caption{V}\label{fig:C}%
    \end{figure}

    \endgroup

    {infinite_sampling}%
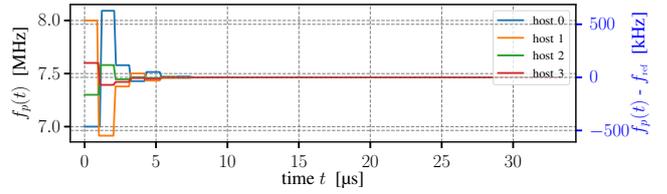
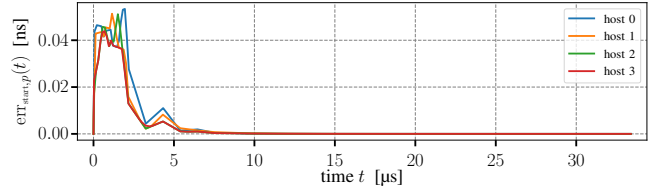
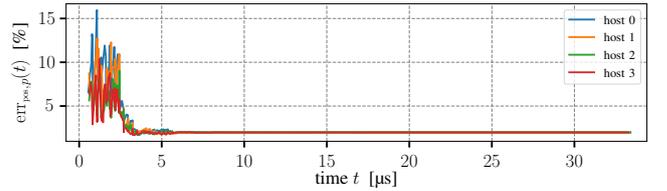
    {Scenario A: Simulation without sampling error}%
    {infinitesampling}

    \begingroup

    \begin{figure}[t]%
        \centering
        \begin{subfigure}{\columnwidth}
            \resizebox{\columnwidth}{!}{\input{\SimulationFiguresPath{}/instantaneous_frequencies.pgf}}%
            \caption{Instantaneous frequency \SimInstantaneousFrequency{p}{t} ($\SimReferenceFrequency = \PlotReferenceFrequencyValue\;\text{\PlotReferenceFrequencyUnit}$)}\label{fig:C_instfreq}
        \end{subfigure}

        \begin{subfigure}{\columnwidth}
            \resizebox{\columnwidth}{!}{\input{\SimulationFiguresPath{}/transmit_deviations.pgf}}
            \caption{Mean absolute differences of \TermBasePulse{} start times}\label{fig:C_txdev}
        \end{subfigure}

        \begin{subfigure}{\columnwidth}
            \resizebox{\columnwidth}{!}{\input{\SimulationFiguresPath{}/relative_position_errors.pgf}}
            \caption{Relative position error}\label{fig:C_poserr}
        \end{subfigure}

        \caption{V}\label{fig:C}%
    \end{figure}

    \endgroup

    {finite_sampling_no_drift_compensation}%
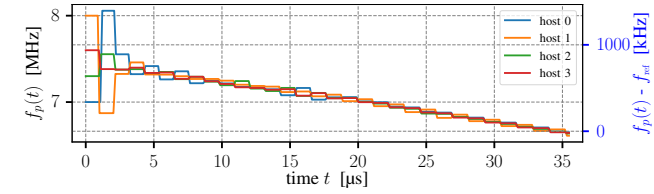
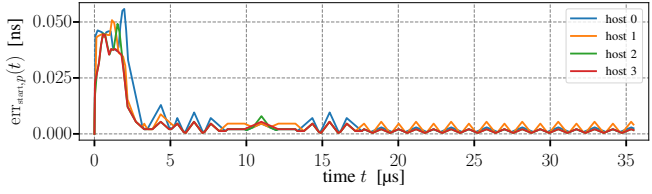
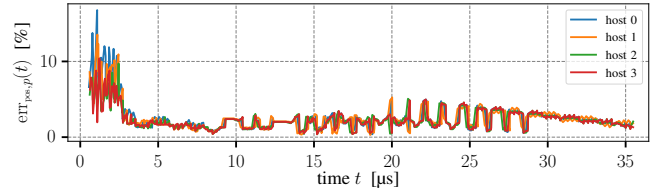
    {Scenario B: Simulation with sampling error and without drift compensation}%
    {finitesamplingnocomp}

    \begingroup

    \begin{figure}[t]%
        \centering
        \begin{subfigure}{\columnwidth}
            \resizebox{\columnwidth}{!}{\input{\SimulationFiguresPath{}/instantaneous_frequencies.pgf}}%
            \caption{Instantaneous frequency \SimInstantaneousFrequency{p}{t} ($\SimReferenceFrequency = \PlotReferenceFrequencyValue\;\text{\PlotReferenceFrequencyUnit}$)}\label{fig:C_instfreq}
        \end{subfigure}

        \begin{subfigure}{\columnwidth}
            \resizebox{\columnwidth}{!}{\input{\SimulationFiguresPath{}/transmit_deviations.pgf}}
            \caption{Mean absolute differences of \TermBasePulse{} start times}\label{fig:C_txdev}
        \end{subfigure}

        \begin{subfigure}{\columnwidth}
            \resizebox{\columnwidth}{!}{\input{\SimulationFiguresPath{}/relative_position_errors.pgf}}
            \caption{Relative position error}\label{fig:C_poserr}
        \end{subfigure}

        \caption{V}\label{fig:C}%
    \end{figure}

    \endgroup

    {finite_sampling_with_continous_drift_compensation}%
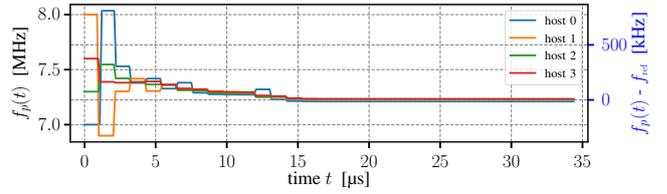
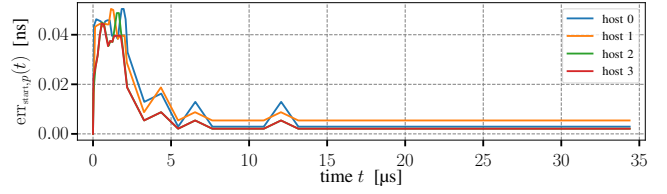
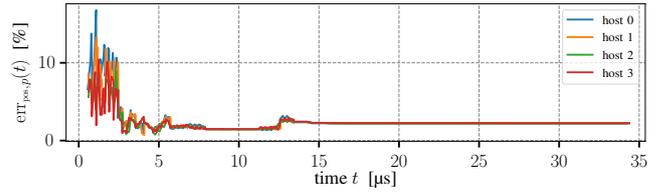
    {Scenario C: Simulation with sampling error and drift compensation}%
    {finitsamplingwithcomp}

\subsection{Results}

Scenario A (\abbildung{infinitesampling}) shows the results for the ideal case without sampling error.
The instantaneous frequencies (\abbildung{infinitesampling_instfreq}) 
and the transmit times (\abbildung{infinitesampling_txdev}) quickly converge to a common and constant value.
Likewise, the localization error (\abbildung{infinitesampling_poserr}) tends to zero in the same amount of time.

\Abbildung{finitesamplingnocomp} depicts the results for scenario B with finite sampling frequency but without drift compensation.
Here, the instantaneous frequencies (\abbildung{finitesamplingnocomp_instfreq}) get closer to each other as well, though not as close as
in the ideal case (scenario A). The main problem is that a joint drift is induced.
As for the start times (\abbildung{finitesamplingnocomp_txdev}), these get closer to each other as well, with some oscillation remaining.
As it can be seen on \abbildung{finitesamplingnocomp_poserr}, the position estimation is unstable under these conditions.
Note from \tabelle{simscenarios} that we used a sampling frequency which is roughly 100 times the initial \TermBasePulse frequency.
Even that did not help to mitigate the negative effects of the induced sampling error.

Finally, scenario C with finite sampling frequency and drift compensation is shown in \abbildung{finitsamplingwithcomp}.
Here we used only a tiny amount of drift compensation as it can be seen \tabelle{simscenarios}.
Still, this small amount was sufficient to combat the frequency drift and to significantly improve the overall synchronization performance.
The instantaneous frequencies (\abbildung{finitsamplingwithcomp_instfreq}) converge to a common and constant value again, 
similar to scenario A,  though the remaining difference is not the same.
Similar results are visible for the transmit times (\abbildung{finitsamplingwithcomp_txdev}). 
The resulting start times are not as close as in scenario A, but significantly better compared to scenario B.
The improved synchronization affects the localization accuracy (\abbildung{finitsamplingwithcomp_poserr}).
Here, the relative error does not approach zero as in scenario A, 
but the position estimates have become stable and much more reliable compared to scenario B.

\section{Conclusion}\label{sec:conclusion}

In this work, we extended the two-stage Kuramoto method for the synchronization of hosts in a wireless network to also support localization capabilities.
We could show that the two-stage Kuramoto method is indeed capable of providing enough information to jointly localize the hosts.
However, we demonstrated that real-world effects like sampling can significantly deteriorate the performance of both, the synchronization and the localization.
To this end, we proposed a simple mechanism to mitigate the negative effects of finite sampling frequencies.
The mitigation improved the frequency and phase synchronization significantly and rendered the localization much more reliable.

\bibliographystyle{IEEEtran}
\bibliography{bibliography}

\end{document}